\documentclass[graphicx, amsmath, bm, docs, 12pt]{iopart}

\usepackage{iopams}
\usepackage{graphicx}
\usepackage{bm}
\usepackage{setstack}
\usepackage{cite}
\makeatletter
\newenvironment{figurehere}
{\def\@captype{figure}} {} \makeatother

\begin{document}

\title[]{Magnetic properties and thermal entanglement on a triangulated Kagom\'{e} lattice}

\author{N S Ananikian$^{1}$, L N Ananikyan$^{1}$, L A Chakhmakhchyan$^{1,2}$ and A N Kocharian$^{1,3}$}

\address{$^1$Yerevan Physics Institute, Alikhanian Br. 2, 0036 Yerevan, Armenia}
\address{$^2$ Yerevan State University, A. Manoogian 1, 0025 Yerevan, Armenia}
\address{$^3$ Department of Physics, California State University, Los Angeles, CA 90032, USA}
\ead{ananik@yerphi.am}
\begin{abstract}
The  magnetic and entanglement thermal (equilibrium) properties in
spin-$1/2$ Ising-Heisenberg model on a triangulated Kagom\'{e}
lattice are analyzed by means of variational mean-field like
treatment based on Gibbs-Bogoliubov inequality. Because of the
separable character of Ising-type exchange interactions between
the Heisenberg trimers the calculation of quantum entanglement
in a self-consistent field can be performed for each of the trimers individually. The
concurrence in terms of three qubit isotropic Heisenberg model in
effective Ising field is non-zero even in the absence of a
magnetic field. The magnetic and entanglement properties exhibit
common (plateau and peak) features observable via (antferromagnetic)
coupling constant and external magnetic field. The critical
temperature for the phase transition and threshold temperature for
concurrence coincide in the case of antiferromagnetic coupling
between qubits. The existence of entangled and
disentangled phases in saturated and frustrated phases is
established.
\end{abstract}

\maketitle

\section{Introduction \label{intr}}
Geometrically frustrated spin systems exhibit a fascinating new
phases of matter, a rich variety of unusual ground states and
thermal properties as a result of zero and finite temperature
phase transitions driven by quantum and thermal fluctuations
respectively~\cite{frust0, frust, frust1, diep, thermal,
thermal1}. The quantum statistical studies of perplexing physics
due to quantum fluctuations, geometric frustration, competing
phases and complex connectivity in electron and spin lattices
still remain a challenging theoretical problem. Some prominent
properties related to the frustrated models, such as transition
from disordered quantum spin liquid into spin insulator with
broken translational symmetry similar to the Mott-Hubbard
insulator at half filling with antiferromagnetic plateau
behavior, seen in magnetization versus magnetic field at
low-temperature, have been recently intensively studied both
experimentally \cite{exper1, exper2, exper3, shastry} and
theoretically \cite{plateaux0, plateaux, plateaux1}. Furthermore,
the (non-bipartite) frustrated local geometry in two and three
dimensions can provide also insight into the magnetism of strongly
correlated electrons in small and large bipartite and
non-bipartite (frustrated) Hubbard lattices also away from half
filling~\cite{nagaoka}.
The efforts aimed at better understanding of the aforementioned phenomena stimulated an intensive search of two-dimensional geometrically frustrated topologies, such as recently fabricated metalo-organic compound piperazinium hexachlorodicuprate \cite{new}, $\mathrm{AgNiO_2}$ triangular magnet \cite{new1}, crystal samples of $\mathrm{KFe_3(OH)_6(SO_4)_2}$ (as an ideal Kagom\'{e} lattice antiferromagnet)
\cite{new2}, alternating Kagom\'{e} and triangular planar layers stacked along the $[111]$ direction
of the pyrochlore lattice \cite{new3}, etc. Besides, studies of inorganic molecular materials with
 paramagnetic metal centers connected in the crystal lattice via superexchange pathways, strongly
 frustrated by their geometric arrangement, has also been implemented \cite{new4}. From this perspective,
the most interesting
geometrically frustrated topologies are the 
magnetic materials in form of two-dimensional isostructural
polymeric coordination compounds $\mathrm{Cu_9X_2(cpa)_6\cdot n
H_2O}$ ($\mathrm{X=F}$, $\mathrm{Cl}$, $\mathrm{Br}$ and
cpa=carboxypentonic acid) \cite{compound2, compound3, compound4}.
The magnetic lattice of these series of compounds consists of
copper ions placed at two non-equivalent positions, which are
shown schematically as open and full circles in figure \ref{TGL}.
$\mathrm{Cu^{2+}}$ ions with a square pyramidal coordination
(a-sites) form equilateral triangles (trimers) which are connected
one to another by $\mathrm{Cu^{2+}}$ ions (monomers) with an
elongated octahedron environment (b-sites) forming the sites of
Kagom\'{e} lattice. This magnetic architecture, which can be
regarded as triangulated (triangles in triangles) Kagom\'{e}
lattice, is currently under active theoretical investigation
\cite{strecka, japan}.

The spin-$\frac{1}{2}$ Ising model on the triangulated Kagom\'{e}
lattice has been exactly solved in \cite{zheng}. However, in its
initial form theory fails to describe the properties of the
aforementioned compound series, since it entirely neglects quantum
fluctuations firmly associated with a quantum nature of the
paramagnetic $\mathrm{Cu^{2+}}$ ions having the lowest possible
quantum spin number 1/2. Further extension to the Ising-Heisenberg
model by accounting for quantum interactions between
$\mathrm{Cu^{2+}}$ ions in $a$-sites (with quantum spin number
1/2) in the limit when monomeric $b$-spins having an exchange of
Ising character, provides much more richer physics and displays
essential features of the copper based coordination compounds
\cite{strecka1, XXZ}. The strong antiferromagnetic coupling has
been assumed for $J_{aa}$ between trimeric $a$ sites, while there
exists a weaker ferromagnetic exchange $J_{ab}$ between the
trimer $a$- and monomer $b$-sites at the ratio $|J_{ab}/J_{aa}|\approx0,025$
\cite{compound1}.

Entanglement is a generic feature of quantum correlations in
systems, which cannot be quantified classically \cite{review,
review1}. It provides a new perspective to understand the quantum
phase transitions (QPTs) and collective phenomena in many-body and
condensed matter physics. This problem, which has been under
scrutiny for nearly two decades, has attracted much attention
recently~\cite{entangle, entangle1, entangle2, entangle3}.

A key observation is that quantum entanglement can play an
important role in proximity to the QPTs, controlled by quantum
fluctuations near quantum critical points~\cite{QPT}. A new line
research points to a connection between the entanglement of a
many-particle system and the existence of the QPTs and scaling
\cite{QPT1, QPT2, alba}.
Thermal entanglement was detected by experimental observations in low
dimensional spin systems formed in the compounds Na$_2$Cu$_5$Si$_4$O$_{14}$ \cite{exp3},  CaMn$_2$Sb$_2$ \cite{exp5},
pyroborate MgMnB$_2$O$_5$ the warwickite MgTiOBO$_3$ \cite{exp4},  KNaMSi$_4$O$_{10}$ (M=Mn, Fe or Cu) \cite{exp1} and metal
carboxylates from magnetic susceptibility  measurements \cite{exp2}.
Entanglement properties for few spins or
electrons can display the local intrinsic features of large
thermodynamic systems and can be suitable for calculations of
basic magnetic quantities and various measures in entanglement
associated with QPTs. The basic features of entanglement in
spin-$\frac{1}{2}$ finite systems are fairly well understood by
now (see e.g. \cite{wang,subrahmanyam}), while the role of local
cluster topology and spin correlations in thermodynamic limit
still remain unanswered.
There are some approximate methods, such
as mean-field like theories based on the Gibbs-Bogoliubov
inequality, that one can invoke to deal with the cases like this, aimed at better
understanding of different physical aspects \cite{meanbook}. This
method can also be  applied for studying thermal entanglement of
many-body systems \cite{mean}. In spite of the method not being
exact, it is still possible to observe regions of criticality
\cite{mean3}.

In the case of the triangulated Kagom\'{e} lattice each $a$-type
trimer interacts with its neighboring trimer through the
Ising-type exchange, i.e. classical interaction, therefore the
states of two neighboring $a$-sublattices become separable
(disentangled) \cite{review}. Thus we can calculate concurrence (a
measure of entanglement \cite{wooters}), which characterizes quantum
features, for each trimer separately in self-consistent field. The
key result of the paper is concentrated on the comparison of
specific (peak and plateau) features in magnetization, susceptibility, specific heat and
thermal entanglement properties in the above mentioned model using
variational mean-field like approximation based on
Gibbs-Bogoliubov inequality. We will demonstrate how the order-disorder phase transition temperature is
relevant to the the threshold temperature for vanishing of
entanglement.

The rest of the paper is organized as follows: in section
\ref{method} we introduce the Ising-Heisenberg model on the
triangulated Kagom\'{e} lattice and provide a solution in
variational mean-field like approximation based on
Gibbs-Bogoliubov inequality. The magnetic properties of the model
are investigated in section \ref{magnetic}. The basic principles for
calculation of concurrence as a measure of entanglement and some
of the results on intrinsic properties are introduced in section
\ref{entanglement}. In section \ref{identity} we present the
comparison of magnetic properties and thermal entanglement. The
concluding remarks are given in section \ref{concl}.

\section{Isotropic Heisenberg
model on triangulated Kagom\'{e} lattice }\label{method}

\subsection{Spin-$\frac{1}{2}$ Ising-Heisenberg model}

We consider the spin-$\frac{1}{2}$ Ising-Heisenberg model on
triangulated Kagom\'{e} lattice (TKL) (figure \ref{TGL}) consisting
of two types of sites ($a$ and $b$). Since the exchange coupling
between $\mathrm{Cu}^{2+}$ ions are almost isotropic, the
application of the $XXX$ Heisenberg model is more appropriate.
There is a strong Heisenberg $J_{aa}$ exchange coupling between
trimeric sites of $a$ type and weaker Ising-type one ($J_{ab}$)
between trimeric $a$ and monomeric $b$ ones. Thus, the Kagom\'{e}
lattice of the Ising spins (monomers) contains inside of each
triangle unit a smaller triangle of the Heisenberg spins (trimer).
The Hamiltonian can be written as follows:

\begin{eqnarray}
{\mathcal{H}}=J_{aa}\sum_{(i,j)}\mathbf{S}^{a}_i\mathbf{S}^{a}_j-J_{ab}\sum_{(k,l)}(S^z)^{a}_k\cdot(S^z)^{b}_l
-H\sum_{i=1}^{\frac{2N}{3}}
3[(S^z)^a_j+\frac{1}{2}(S^z)^b_j], \label{1}
\end{eqnarray}
where  $\mathbf{S}^a=\{S^a_x, S^a_y, S^a_z\}$ is the Heisenberg
spin-$\frac{1}{2}$ operator, $S^b$ is the Ising spin. $J_{aa}>0$
corresponds to antiferromagnetic Heisenberg coupling and $J_{ab}>0$ to
ferromagnetic Ising-Heisenberg one. The first two summations run over $a-a$ and
$a-b$ nearest neighbors respectively and the last sum incorporates
the effect of uniform magnetic field (we have assumed that the
total number of sites is $3N$).

\begin{figurehere}
\includegraphics[width=10cm]{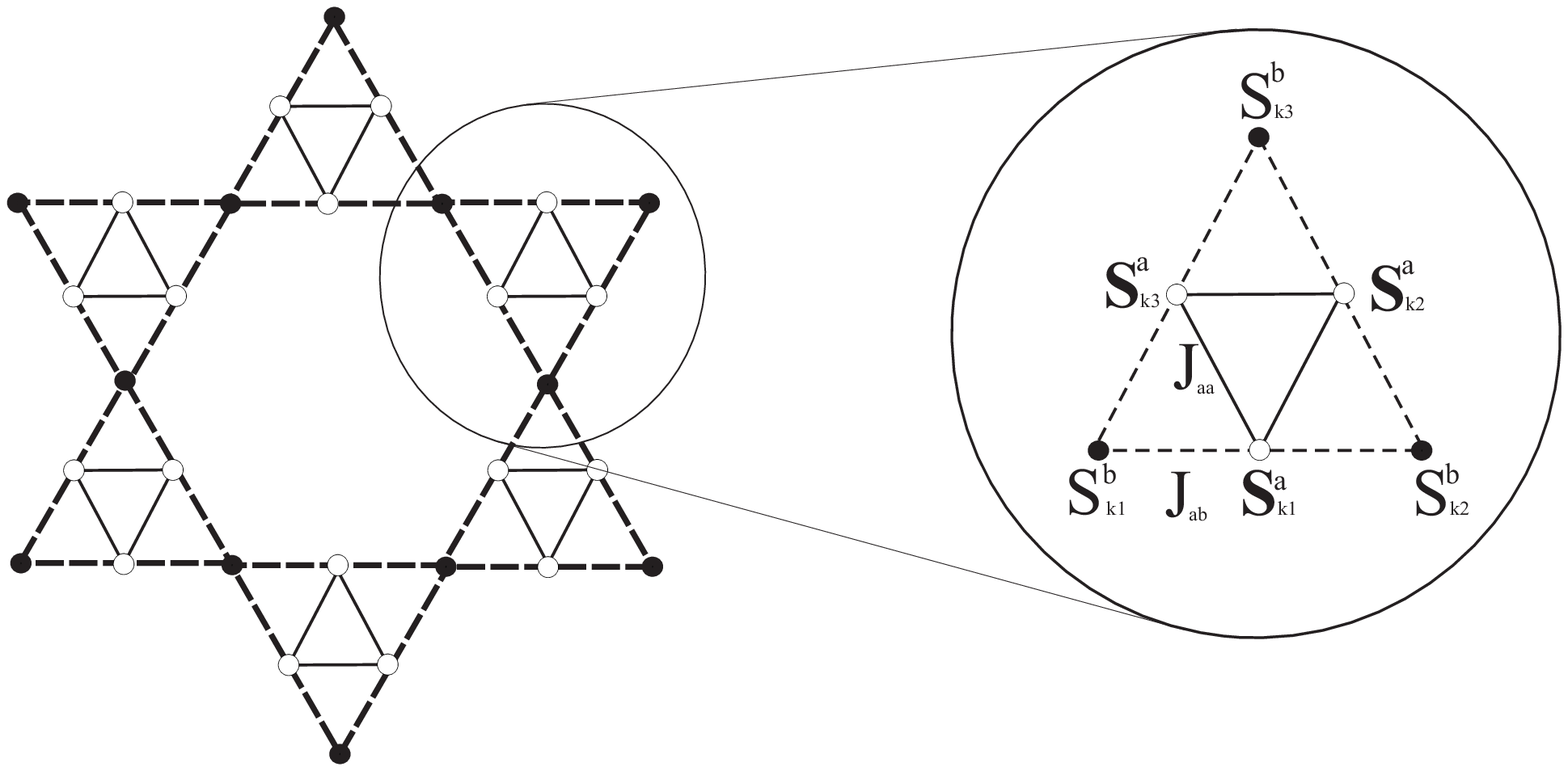}
\caption{ \small {A cross-section of TKL structure. Solid lines
represent the intra-trimer Heisenberg interactions $J_{aa}$, while
the broken ones label monomer-trimer Ising interactions $J_{ab}$.
The circle marks $k$-th cluster (Heisenberg trimer).
$\mathbf{S}^a_{k_i}$ presents the Heisenberg and $S^b_{k_i}$ the
Ising spins. \label{TGL}}}
\end{figurehere}

\subsection{Basic mean-field formalism}
Here we apply the variational mean-field like treatment based on
Gibbs-Bogoliubov inequality~\cite{gibbs} to solve the Hamiltonian
(\ref{1}). This implies that
the free energy (Helmholtz potential) of system is
\begin{equation}
F\leq F_0+\langle{\mathcal{H}}-{\mathcal{H}_0}\rangle_0, \label{2}
\end{equation}
where $\mathcal{H}$ is the real Hamiltonian which describes the
system and $\mathcal{H}_0$ is the trial one. $F$ and $F_0$ are free
energies corresponding to $\mathcal{H}$ and $\mathcal{H}_0$
respectively and $\langle...\rangle_0$ denotes the thermal average
over the ensemble defined by $\mathcal{H}_0$. Following
\cite{strecka} we introduce the trial Hamiltonian in the following
form: \numparts
\begin{eqnarray}\label{3a}
\mathcal{H}_0=&\sum_{k \in trimers}\mathcal{H}_{c_0},\\
\mathcal{H}_{c_0}=&\lambda_{aa}\left(\mathbf{S}^{a}_{k_1}\mathbf{S}^{a}_{k_2}+
\mathbf{S}^{a}_{k_2}\mathbf{S}^{a}_{k_3}+\mathbf{S}^{a}_{k_1}\mathbf{S}^{a}_{k_3}\right)-\sum_{i=1}^{3}
\left[\gamma_a(S^z)^a_{k_i}+\frac{\gamma_b}{2}(S^z)^b_{k_i}\right].
\label{3b}
\end{eqnarray}
\endnumparts
In this Hamiltonian the stronger quantum Heisenberg
antiferromagnetic interactions between $a$-sites are treated
exactly, while the weaker Ising-type ones between $a$- and $b$-sites
are replaced by self-consistent (effective) fields of two types:
$\gamma_a$ and $\gamma_b$. The variational parameters $\gamma_a$,
$\gamma_b$ and $\lambda_{aa}$ can be found from the Bogoliubov
inequality after minimizing the RHS of (\ref{2}). Thus our focus is
on the cluster depicted in figure~\ref{TGL}.
Each of the $b$-type spins belongs to two of such clusters simultaneously. Hence only the half of each $b$-spin belongs to one cluster and therefore there are $3\cdot1/2$ b-type spins in it. Consequently the total number of
($a-$ and $b-$) spins in the cluster will be $3+3/2=9/2$ (there are $3$ $a$-type spins in the triangle).
Inequality (\ref{2}) can be
rewritten for the described cluster:
\begin{equation}
f\leq f_0+\langle{\mathcal{H}_c}-\mathcal{H}_{c_0}\rangle_0,
\label{4}
\end{equation}
where $\mathcal{H}_c$ is the real and $\mathcal{H}_{c_0}$ the
trial Hamiltonian of the cluster, $f$ and $f_0$ free energies of
the cluster defined by $\mathcal{H}_c$ and $\mathcal{H}_{c_0}$
respectively. Using the fact that in terms of (\ref{3b})
$\mathbf{S}^{a}$ and $\mathbf{S}^{b}$ are statistically
independent, one obtains $\langle
\mathbf{S}^{a}\cdot\mathbf{S}^{b}\rangle_0=\langle\mathbf{S}^{a}\rangle_0\cdot\langle\mathbf{S}^{b}\rangle_0$.
Besides, taking into account that
$\langle{(S^z)^a}\rangle_0=m_a$ (single $a$-site magnetization),
$\langle{(S^z)^b}\rangle_0=m_b$ (single $b$-site magnetization),
we obtain the following expression:
\begin{eqnarray}
f\leq &&f_0+(J_{aa}-\lambda_{aa})\langle
\mathbf{S}^{a}_{k_1}\mathbf{S}^{a}_{k_2}
+\mathbf{S}^{a}_{k_2}\mathbf{S}^{a}_{k_3}+\mathbf{S}^{a}_{k_1}\mathbf{S}^{a}_{k_3}\rangle_0 \nonumber \\
&&-6J_{ab}m_a m_b-3H m_a-\frac{3H m_b}{2}+3\gamma_a
m_a+\frac{3\gamma_b m_b}{2}. \label{5}
\end{eqnarray}
Now, by minimizing the right-hand side of inequality (\ref{5}) with respect to $\gamma_a$, $\gamma_b$
and $\lambda_{aa}$ and using $\frac{\partial
f_0}{\partial \gamma_a}=-3m_a$, $\frac{\partial f_0}{\partial
\gamma_b}=-3/2m_b$, $\frac{\partial f_0}{\partial
\lambda_{aa}}=\langle \mathbf{S}^{a}_{k_1}\mathbf{S}^{a}_{k_2}+
\mathbf{S}^{a}_{k_2}\mathbf{S}^{a}_{k_3}+\mathbf{S}^{a}_{k_1}\mathbf{S}^{a}_{k_3}\rangle_0$,
we determine the variational parameters in the form:
$\lambda_{aa}=J_{aa}$, $\gamma_a=2J_{ab}m_b+H$,
$\gamma_b=4J_{ab}m_a+H$.
Parameters $\gamma_a$ and $\gamma_b$, which have a meaning of a
magnetic field, are interconnected, which is the consequence of
its` apparent self-consistency.
The Hamiltonian $\mathcal{H}_0$ was chosen to be exactly solved.
One finds that $\mathcal{H}_{c_0}$ can be divided into two parts
corresponding to $a$- and $b$-type variables:
\begin{eqnarray} \nonumber
\mathcal{H}_{c_0}=&&\left[\lambda_{aa}\{\mathbf{S}^{a}_{k_1}\mathbf{S}^{a}_{k_2}+
\mathbf{S}^{a}_{k_2}\mathbf{S}^{a}_{k_3}+\mathbf{S}^{a}_{k_1}\mathbf{S}^{a}_{k_3}\}-\sum_{i=1}^{3}
\gamma_a(S^z)^a_{k_i}\right]-
\sum_{i=1}^{3}\frac{\gamma_b}{2}(S^z)^b_{k_i}\nonumber \\
&&=\mathcal{H}_{c_0}^a+\sum_{i=1}^{3}(\mathcal{H}_{c_0}^b)^i
\label{6}.
\end{eqnarray}
Each of Hamiltonians $\mathcal{H}_{c_0}^a$ and $(\mathcal{H}_{c_0}^b)^i$ can be solved separately (the variables have been separated).
The eigenvalues of $\mathcal{H}_{c_0}^a$ are:
\begin{eqnarray} \nonumber
&E_1=\frac{3}{4} \left(\lambda_{aa} +2 \gamma _a\right); \quad
E_2=E_3=\frac{1}{4} \left(-3 \lambda_{aa}+2 \gamma _a\right);\\
&E_4=\frac{1}{4} \left(3 \lambda_{aa} +2 \gamma _a\right); \quad E_5=E_6=\frac{1}{4} \left(-3 \lambda_{aa} -2 \gamma _a\right);\\
\nonumber &E_7=\frac{1}{4} \left(3 \lambda_{aa} -2 \gamma _a\right);
\quad E_8=\frac{3}{4} \left(\lambda_{aa} -2 \gamma
_a\right) \label{8}
\end{eqnarray}
and the corresponding eigenvectors given by
\begin{eqnarray}\nonumber
&|\psi_1\rangle=|000\rangle\\ \nonumber
&|\psi_2\rangle=\frac{1}{\sqrt{3}}\left(q|001\rangle+q^2|010\rangle+|100\rangle\right)&\\
\nonumber
&|\psi_3\rangle=\frac{1}{\sqrt{3}}\left(q^2|001\rangle+q|010\rangle+|100\rangle\right)&\\
\nonumber
&|\psi_4\rangle=\frac{1}{\sqrt{3}}\left(|001\rangle+|010\rangle+|100\rangle\right)&\\
&|\psi_5\rangle=\frac{1}{\sqrt{3}}\left(q|110\rangle+q^2|101\rangle+|011\rangle\right)&\\
\nonumber
&|\psi_6\rangle=\frac{1}{\sqrt{3}}\left(q^2|110\rangle+q|101\rangle+|011\rangle\right)&\\
\nonumber
&|\psi_7\rangle=\frac{1}{\sqrt{3}}\left(|110\rangle+|101\rangle+|011\rangle\right)&\\
\nonumber &|\psi_8\rangle=|111\rangle,  \nonumber
\end{eqnarray}
where $q=e^{i2\pi/3}$ (these eigenvectors should be also the
eigenstates of cyclic shift operator $P$ with eigenvalues $1$, $q$
and $q^2$, satisfying $q^2+q+1=0$).

The partition function $Z_{0_a}$ of the trimer in mean-field approximation is:
\begin{eqnarray}
Z_{0_a}=\sum_{k=1}^8\exp(-E_k/T)=e^{-\frac{3{\lambda_{aa}}}{4 T}}&\left[\cosh \left(\frac{3 \gamma _a}{2 T}\right)+\right.
\\ \nonumber
&\left.2 e^{\frac{3{\lambda_{aa}}}{2 T}} \cosh \left(\frac{\gamma _a}{2T}\right)+\cosh\left(\frac{\gamma _a}{2 T}\right)\right]. \label{80}
\end{eqnarray}
Consequently the free energy of $a$-triangle will be:
\begin{eqnarray}
f_{0_a}=-T\ln Z_{0_a}=\frac{3{\lambda_{aa}}}{4}-T\ln&&\left[\cosh \left(\frac{3 \gamma _a}{2 T}\right)+2 e^{\frac{3{\lambda_{aa}}}{2 T}}\cosh \left(\frac{\gamma _a}{2T}\right)+\right.\\ \nonumber
&&\left.\cosh\left(\frac{\gamma _a}{2 T}\right)\right]. \label{123}
\end{eqnarray}
Since the $(\mathcal{H}_{c_0}^b)^i$ describes only half a particle (b-type spin), the Hamiltonian of one b-type spin will be $2\cdot(\mathcal{H}_{c_0}^b)^i$. Hence, following the technique described above one finds the partition function $Z_{0_b}$ and free energy $f_{0_b}$  of a b-type spin in the adopted approximation:
\numparts
\begin{eqnarray}\label{123a}
Z_{0_b}=2\cosh\left(\frac{\gamma _b}{2 T}\right),\\
f_{0_b}=-T\ln\left[2\cosh\left(\frac{\gamma _b}{2 T}\right)\right].
\label{123b}
\end{eqnarray}
\endnumparts
As already mentioned, in terms of the trial Hamiltonian $a$- and $b$-type spins are statistically independent [see equations (3) and (\ref{6})]. Besides, $b-$ type spins do not interact with each other. Therefore the partition function $f_0$ of the cluster in mean-field approximation reads:
\begin{equation}
f_0 =f_{0_a}+\frac{3}{2}\cdot f_{0_b} \label{7.33}.
\end{equation}
Consequently the free energy of the cluster $f_{GB}$ in the mean-field approximation based on the Gibbs-Bogoliubov inequality will be:
\begin{eqnarray}
f_{GB}=&&f_0+\langle{\mathcal{H}_c}-\mathcal{H}_{c_0}\rangle_0= \frac{3\lambda_{aa}}{4}+6 J_{ab} m_a m_b-T \left[\ln
\left\{4 e^{\frac{3
   J_{ab}}{2 T}} \cosh \left(\frac{\gamma _a}{2 T}\right)\right.\right. \nonumber \\
   && \left.+2 \cosh\left(\frac{\gamma _a}{2 T}\right)+2 \cosh \left(\frac{3 \gamma _a}{2
   T}\right)\right\}  \left.+\frac{3}{2} \ln \left\{2 \cosh \left(\frac{\gamma
   _b}{2 T}\right)\right\}\right]. \label{9}
\end{eqnarray}
In equation (\ref{9}) we have used the values of variational parameters $\gamma_a$, $\gamma_b$ and $\lambda_{aa}$.
Besides, due to the fact that there are $9/2$ spins in the cluster and therefore totally $2N/3$ clusters ($F_{GB}=2N/3\cdot f_{GB}$), we
obtain:
\begin{eqnarray}
\frac{F_{GB}}{N}=&& \frac{\lambda_{aa}}{2}+4 J_{ab} m_a m_b-2T
\left[\frac{1}{3}\ln \left\{4 e^{\frac{3
   J_{ab}}{2 T}} \cosh \left(\frac{\gamma _a}{2 T}\right)\right.\right. \nonumber \\
   &&\left.+2 \cosh\left(\frac{\gamma _a}{2 T}\right)+2 \cosh \left(\frac{3 \gamma _a}{2
   T}\right)\right\} \left.+\frac{1}{2} \ln \left\{2 \cosh \left(\frac{\gamma
   _b}{2 T}\right)\right\}\right].  \label{10}
\end{eqnarray}
As for defined above $a$- and $b$-single site magnetizations we
obtain: \numparts
\begin{eqnarray}\label{8a}
m_a=&-\frac{1}{3}\frac{\partial
f_{0_a}}{\partial \gamma_a}=\frac{1}{6}\frac{3 \sinh \left(\frac{3 \gamma _a}{2
T}\right)+2 e^{\frac{3 {\lambda_{aa}}}{2 T}} \sinh
\left(\frac{\gamma _a}{2 T}\right)+\sinh \left(\frac{\gamma _a}{2
T}\right)}{\cosh \left(\frac{3 \gamma _a}{2 T}\right)+2 e^{\frac{3
{\lambda_{aa}}}{2 T}} \cosh \left(\frac{\gamma _a}{2
T}\right)+\cosh
\left(\frac{\gamma _a}{2 T}\right)},\\
m_b=&-\frac{\partial
f_{0_b}}{\partial \gamma_b}=\frac{1}{2} \tanh \left(\frac{\gamma _b}{2 T}\right).
\label{8b}
\end{eqnarray}
\endnumparts

Notwithstanding of simplicity and the fact that the effective
(self-consistent) field in zero magnetic field ($H=0$) overestimates
ferromagnetic correlations, it is still particularly useful for
detection of spontaneous breaking SU(2) symmetry and possible
temperature driven transitions in the frustrated spin systems (see
section \ref{mag_T_H}). However, we also find that the strong
quantum fluctuations, existing in the isotropic Heisenberg model
in the absence of ferromagnetic type Ising term at $H=0$ can
restore the broken symmetry by providing stability to disordered
spin-$1/2$ liquid state in frustrated geometries. Moreover, in
general the presence of magnetic field ($H\neq 0$) suppresses the
spin fluctuations and makes the self-consistent results more
reliable and accurate \cite{yang}. Therefore, the equations
(7), (8) and (13)-(15) with magnetic field are quite sufficient for
understanding some intrinsic relationships between magnetic and
entanglement properties \cite{wang} that naturally emerge in the
Ising-Heisenberg model when one is complying with the variational
mean-field like procedure.

\section{Magnetic properties \label{magnetic}}

\subsection{Magnetization \label{mag_T_H}}

The results of the previous section can be used for investigation
of the magnetic properties of the model. Here we are interested in
the sublattice $a$ properties, which, however, depend on
parameters describing $b$-type spins. It is convenient to
introduce a new (ratio) parameter $\alpha=J_{ab}/J_{aa}$. The
magnetization curves can be found by solving numerically the
transcendental equations (\ref{8a}) and (\ref{8b}). The magnetic field
dependence of the magnetization per atom is plotted in figure
\ref{mag} at $\alpha=0.025$. We find that in the absence of
magnetic field, the
ordered ferromagnetic phase with spontaneous magnetization
per
site $m_{a}$
is a stable ground state for all $|J_{ab}/J_{aa}|$ in spite of the
high geometric frustration caused by the non-bipartite structure
and antiferromagnetic intra-trimer interaction.

\begin{center}
\begin{figurehere}
\begin{tabular}{ c c }
\small(a)  &  \small(b) \\
\begin{figurehere}
\includegraphics[width=7cm]{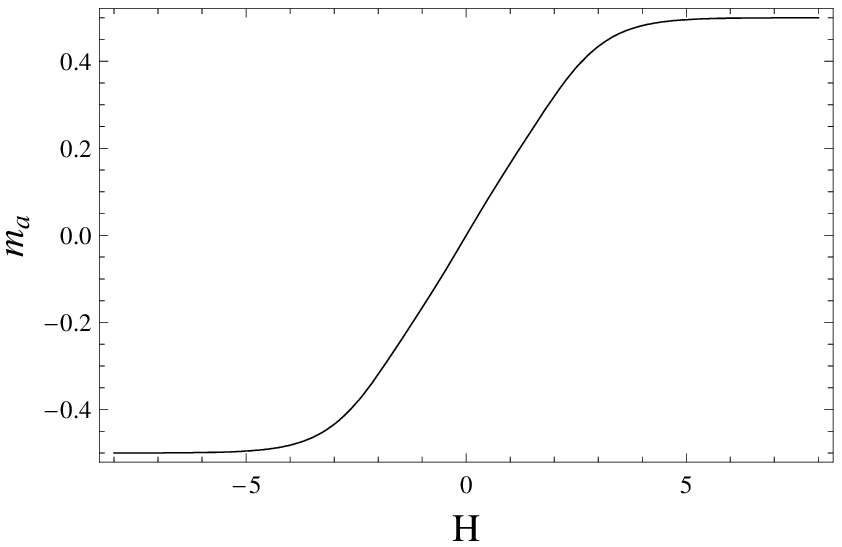}
\end{figurehere} &
\begin{figurehere}
\includegraphics[width=7cm]{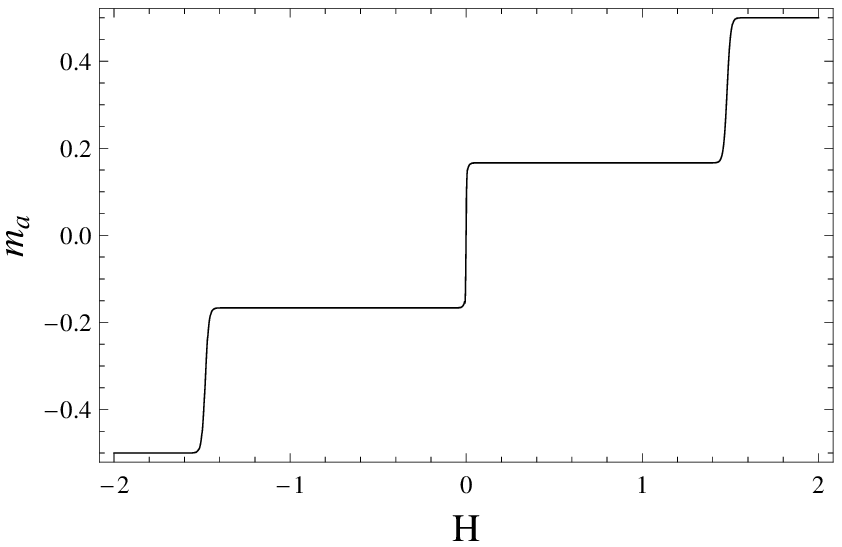}
\end{figurehere}  \\
\end{tabular}
\caption {\small{Single $a$-site magnetization $m_{a}$ versus
external magnetic field $H$ for (a) $T=0.7$ and (b) $T=0.01$ for
$\alpha=0.025$, $J_{aa}=1$. The magnetic field and temperature are
in relative units of $k_B=1$.\label{mag}}}
\end{figurehere}
\end{center}

At relatively high temperatures the magnetization curve in
figure~\ref{mag}(a) shows a monotonic behavior versus magnetic field
with a full saturation at strong magnetic field. Upon decreasing the
temperature a new partially saturated phase emerges in form of the
(spin) plateaus shown in figure~\ref{mag}(b), which can be
associated with staggered magnetization or short range
antiferromagnetism (AF) in frustrated Kagom\'{e} geometry. Indeed,
the appearance of plateau in magnetization curve at $m_a=1/6$ can be
explained as stability of trimeric $a$ sites in
$\uparrow\uparrow\downarrow$ configuration. Thus, at rather low
temperatures, the magnetization shows the finite leap across a
plateau at $m_a=1/6$ at infinitesimal magnetic field and below the
critical field for full saturation by flipping a down spin.

In figure~\ref{mag_T} we also show the temperature dependence of the
magnetization in equilibrium. As one can see from
figure~\ref{mag_T}(a), in the absence of the external magnetic field
the magnetization tends gradually to zero near the second-order
transition temperature
 $T_c$ between ordered ($m_a\neq0$) and disordered ($m_a=0$) phases.
Hence, the
magnetization $m_a$ can be expanded into series near the critical
temperature of second-order phase transition point:
\begin{equation}
m_a = am_a + bm_a^3 + cm_a^5 +... \label{13}
\end{equation}
The critical temperature $T_c$ corresponding to the second-order
phase transition can be found from the condition $a=1$, $b<0$. In
particular, for the case $J_{aa}=1$ and $\alpha=0.025$,
$T_c=0.0102062$ (in relative units).
At zero temperature in the absence of
field we find unsaturated spontaneous ferromagnetism with $m_a=1/6$
as a stable ground state.

\begin{center}
\begin{figurehere}
\begin{tabular}{ c c }
\small(a)  &  \small(b) \\
\begin{figurehere}
\includegraphics[width=7cm]{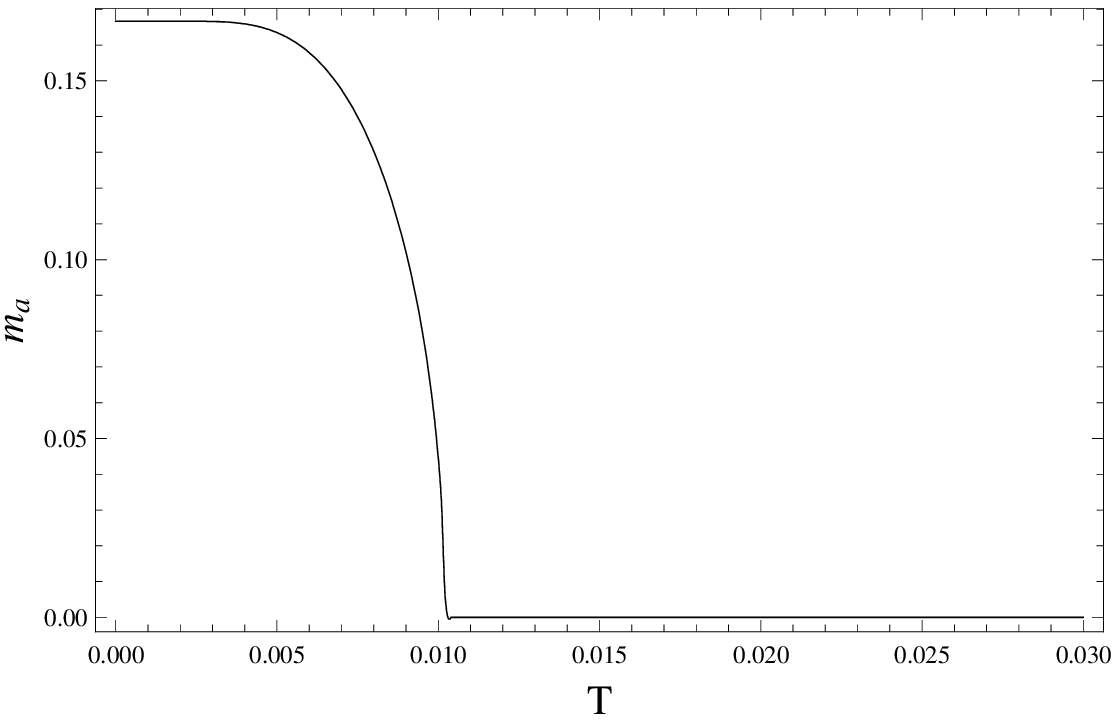}
\end{figurehere} &
\begin{figurehere}
\includegraphics[width=7cm]{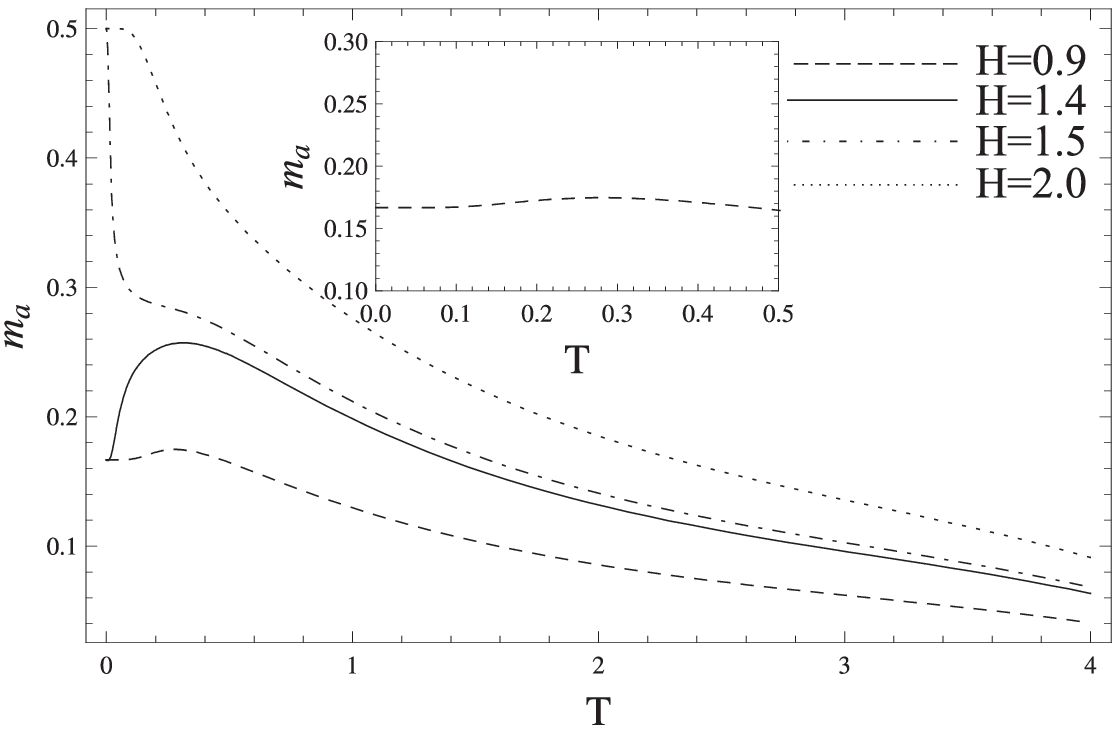}
\end{figurehere}  \\
\end{tabular}
\caption {\small{Single $a$-site magnetization $m_{a}$ per atom
versus temperature $T$ for $J_{aa}=1$, $\alpha=0.025$ and (a) $H=0$
and (b) different non-zero values of $H$ (the inset shows details
for the case $H=0.9$ at low temperatures). \label{mag_T}}}
\end{figurehere}
\end{center}

The magnetization in equilibrium
as a function of temperature $T$ at non-zero magnetic field
\cite{mag_t} is plotted in figure \ref{mag_T}(b). There are two
distinct magnetic field regimes corresponding to $m_a=1/6$ and
saturated zero-temperature magnetization, $m_a=1/2$. While fixed
magnetic field $H$ is less than the saturation magnetic field
value, we deal with $m_a=1/6$ regime. If we continue increasing
the value of $H$, at the saturation magnetic field the
magnetization jumps into $m_a=1/2$ regime [$H=1.5,$ $2.0$ in
figure \ref{mag_T}(b)]. There can be also seen a short plateau at
$m_a=1/6$ in the temperature dependence [$H=0.9$ in figure
\ref{mag_T}(b)]. The peaks in the case of low magnetic fields
arise due to the frustration effects.

\subsection{Susceptibility \label{susc}}

We define the magnetic susceptibility $\chi_a$ as

\begin{equation}
\chi_a=\frac{\partial m_a}{\partial H}. \label{13}
\end{equation}
First we examine the zero-field susceptibility, which is
introduced as follows:
\begin{equation}
\chi_{a_0}=\left.\frac{\partial m_a}{\partial H}\right|_{H=0}.
\label{14}
\end{equation}
The temperature dependence of $\chi_{a_0}$ in equilibrium is plotted in figure \ref{chi_0}.
\begin{figurehere}
\begin{center}
\includegraphics[width=8cm]{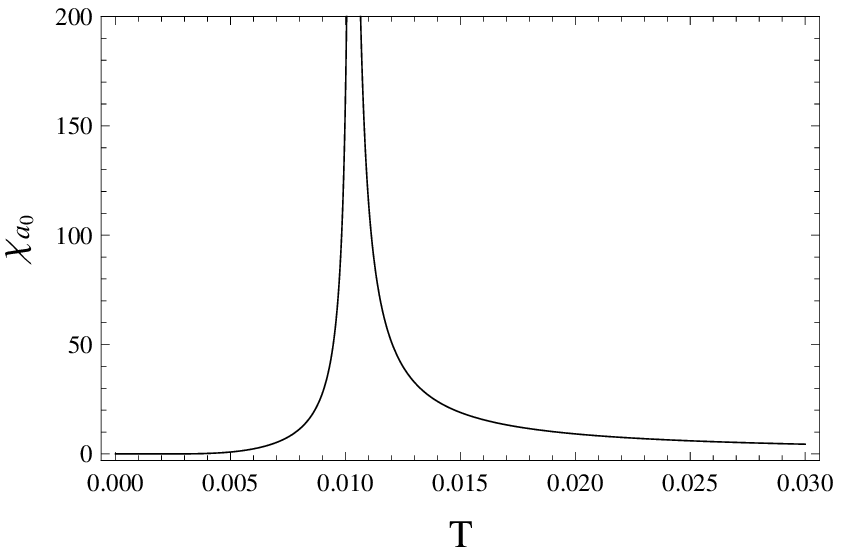}
\caption { \small {Zero-field susceptibility $\chi_{a_0}$ versus
temperature $T$ for $J_{aa}=1$ and $\alpha=0.025$.}} \label{chi_0}
\end{center}
\end{figurehere}
The zero-field susceptibility $\chi_{a_0}$ diverges at the critical
temperature $T_c$ which is a signature of the second order phase
transition discussed earlier (see section \ref{mag_T_H}).
The temperature
dependence of susceptibility $\chi_a$ at $H\neq 0$ \cite{honecker}
is presented in figure \ref{chi_H_T}(a).
\begin{center}
\begin{figurehere}
\begin{tabular}{ c c }
\small(a)  &  \small(b) \\
\begin{figurehere}
\includegraphics[width=7cm]{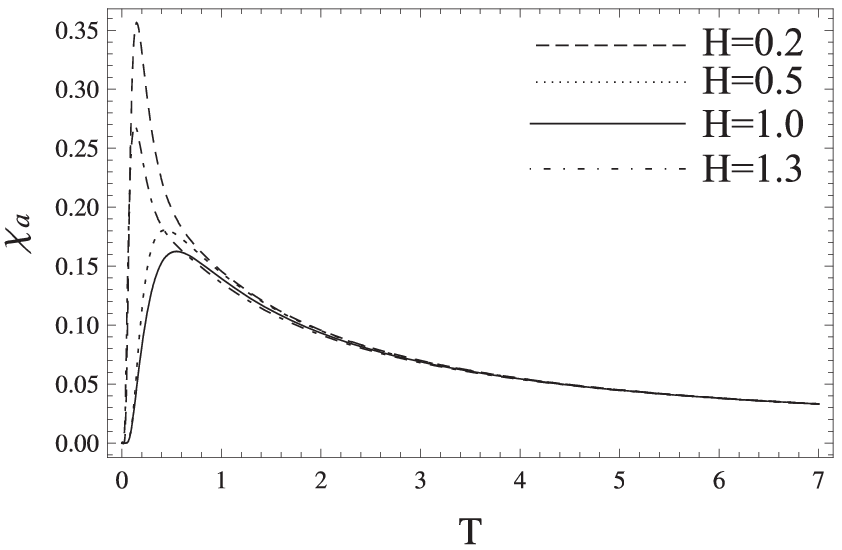}
\end{figurehere} &
\begin{figurehere}
\includegraphics[width=7cm]{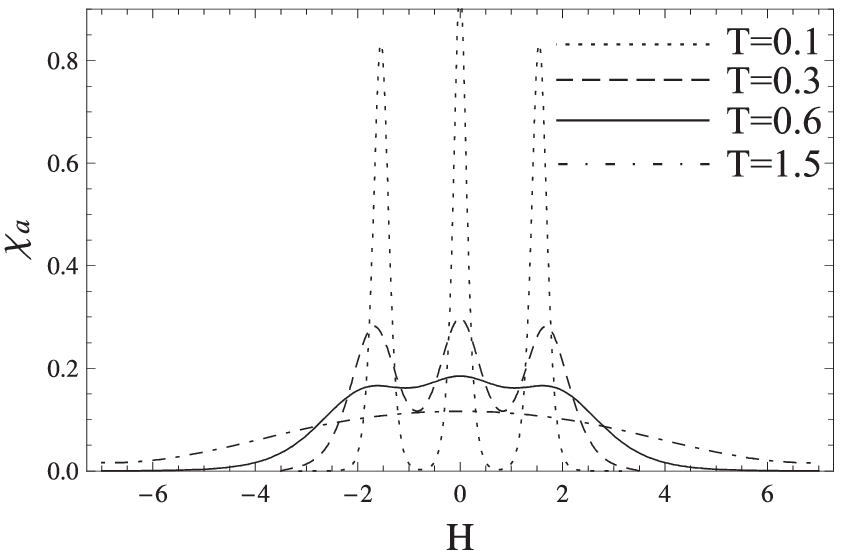}
\end{figurehere}  \\
\end{tabular}
\caption {\small{Susceptibility $\chi_a$ versus (a)
temperature $T$
at different $H$ and (b) magnetic field $H$ at different $T$ for
$J_{aa}=1$, $\alpha=0.025$. \label{chi_H_T}}}
\end{figurehere}
\end{center}

The temperature dependence of magnetic susceptibility observed in
\cite{experim} for $\mathrm{[Ni(H_2L^2)]_4[Cr(CN)_6]_5OH\cdot
15H_2O}$ compounds resembles the result shown in figure
\ref{chi_0}.

Notice, that at high temperatures
the external field dependence of the magnetic susceptibility
[figure \ref{chi_H_T}(b)] exhibits one peak. With
decreasing temperature, two symmetric peaks begin to arise, which
correspond to the formation of the incipient (magnetization)
plateau at $m_a=1/6$. With decreasing further temperature
the peaks become sharper and bigger in their
magnitude.

\subsection{Specific heat \label{heat}}
The internal energy $u$ and the specific heat $c(T)$
per cluster site are, respectively, determined as
\begin{eqnarray}
u=-T^2\frac{\partial}{\partial T}\left(F_{GB}/3NT\right)\\
\label{14} c(T)=\frac{\partial u}{\partial T}=-\frac{T}{3N}\frac{\partial^2
F_{GB}}{\partial T^2},\label{15}
\end{eqnarray}
$F_{GB}$ taken from (\ref{10}).
\begin{figurehere}
\begin{center}
\includegraphics[width=8cm]{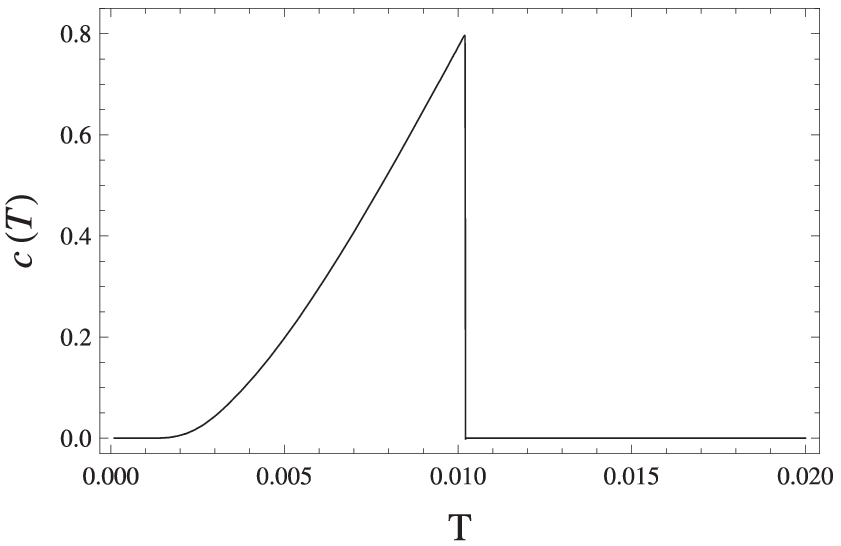}
\caption { \small { The zero-field specific $c(T)$ heat versus
temperature $T$ for $J_{aa}=1$, $\alpha=0.025$.}} \label{spec_0}
\end{center}
\end{figurehere}

The behavior of the specific heat in
equilibrium at
the absence of the external magnetic field ($H=0$) is shown in
figure \ref{spec_0}. In this plot one can
find the presence
of second order phase transition: at the same temperature $T_c$,
described in last two subsections, the specific heat has
discontinuity.

\begin{figurehere}
\begin{center}
\includegraphics[width=8cm]{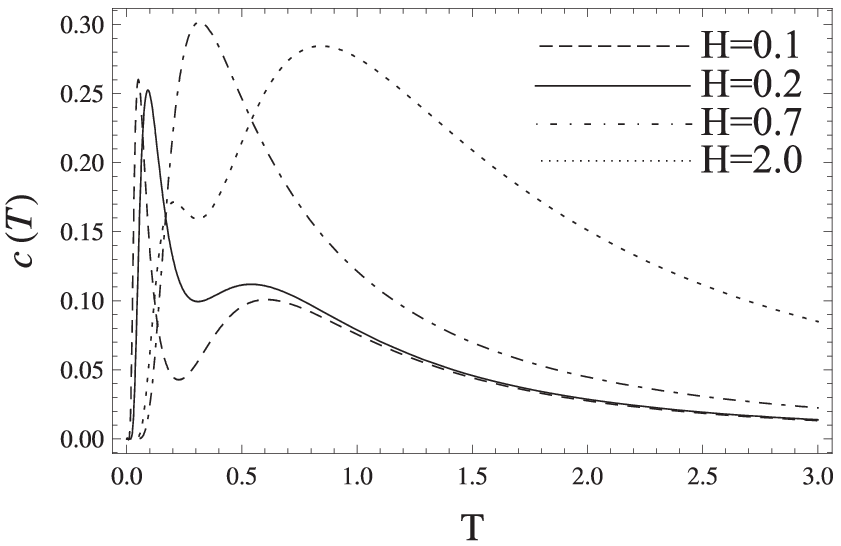}
\caption { \small {Specific heat $c(T)$ versus temperature $T$ for
$J_{aa}=1$, $\alpha=0.025$ and different $H$ values.}}
\label{spec_T}
\end{center}
\end{figurehere}

The temperature dependence of the specific heat at non-zero
magnetic field is shown in figure \ref{spec_T}.
Notwithstanding the observation of one peak in the $c(T)$ at $H=0$
the temperature dependence of specific heat at $H\neq 0$ exhibits
two-peak behavior peculiar to one and quasi one dimensional
systems as one can find in \cite{specheat, twopick}. A double-peak
structure in the specific heat manifests the existence of two
energy scales in the system as a result of two competing orders
\cite{abouie}. Upon increasing
the external magnetic field the peak moves to
higher-temperature region and, at the same time, decreases in
amplitude. At higher values of $H$ close to the transition from
$m_a=1/6$ to saturated $m_a=1/2$ state the second broad peak
gradually increases.

In section \ref{identity} we further discuss magnetic properties by
comparison with thermal entanglement.

\section{Concurrence and thermal entanglement \label{entanglement}}

The mean-field like treatment of (\ref{1}) transforms many-body
system to reduced "single" cluster study in a self-consistent field
where quantum interactions exist. This allows to study, in
particular, (local) thermal entanglement properties of
$a$-sublattice in terms of three-qubit $XXX$ Heisenberg model in
effective magnetic field $\gamma_a$, which carries the main
properties of the system. Besides, because of the self-consistency
and interconnection of the fields $\gamma_a$ and $\gamma_b$ the
effective $\gamma_b$ field have an impact on the concurrence, too.
We study \textit{concurrence} $C(\rho)$,
to quantify
pairwise
entanglement \cite{wooters}, defined as
\begin{equation}
C(\rho)=max\{\lambda_1-\lambda_2-\lambda_3-\lambda_4, 0\}, \label{16}
\end{equation}
where $\lambda_i$ are the square roots of the eigenvalues of the
corresponding operator for the density matrix
\begin{equation}
\tilde{\rho}=\rho_{12}(\sigma_1^y\otimes\sigma_2^y)\rho_{12}^*(\sigma_1^y\otimes\sigma_2^y)
\label{17}
\end{equation}
in descending order. Since we consider pairwise entanglement, we
use reduced density matrix $\rho_{12}=\mathrm{Tr_3}\rho$. Before
introducing the calculations and discussion we would like to
emphasize the fact which was already discussed in section \ref{intr}:
the states of two neighboring $a$-type trimers are separable
(disentangled). Hence we can calculate
the concurrence for each of them on cluster level individually in
effective magnetic field. In our case the density matrix has the
following form
\begin{equation}
\rho=\frac{1}{Z_0{_a}}\sum_{k=1}^8\exp(-E_k/T)|\psi_k\rangle
\langle\psi_k|, \label{7.1}
\end{equation}
$E_k$, $|\psi_k\rangle$ and $Z_0{_a}$ are taken from equations (7), (8) and (9) respectively.
The construction process of matrix (\ref{17}) does not depend
whether $\gamma_a$ is effective or real magnetic field,
although the presence of effective field $\gamma_a$ plays crucial
role for the self-consistent solution.
Here we skip the specific details and provide the result of final
calculations of the matrix $\rho_{12}$, taking into account that the Hamiltonian $\mathcal{H}_{c_0}$ is translationary invariant with a symmetry $[S_z, \mathcal{H}_{c_0}]=0$ ($S_z=\sum_{k=1}^3(S_z)^a_{k_i}$). Hence \cite{matrix}:
\begin{equation}
\rho_{12}=\left(
\begin{array}{llll}
 u & 0 & 0 & 0 \\
 0 & w & y & 0 \\
 0 & y^* & w & 0 \\
 0 & 0 & 0 & v
\end{array}
\right),\label {19}
\end{equation}
where
\begin{eqnarray}
&u=\frac{1}{3} e^{\frac{2 \gamma _a-3 \lambda _{aa}}{4 T}}
\left(1+3
   e^{\frac{\gamma _a}{T}}+2 e^{\frac{3 \lambda _{aa}}{2
   T}}\right)\\
&v=\frac{1}{3} e^{-\frac{3 \left(2 \gamma _a+\lambda
_{aa}\right)}{4
   T}} \left(3+e^{\frac{\gamma _a}{T}}+2 e^{\frac{2 \gamma _a+3 \lambda
   _{aa}}{2 T}}\right)\\ \label{20}
&w=\frac{1}{3} e^{-\frac{2 \gamma _a+3 \lambda _{aa}}{4 T}}
   \left(1+e^{\frac{\gamma _a}{T}}\right) \left(1+2 e^{\frac{3 \lambda
   _{aa}}{2 T}}\right)\\
&y=-\frac{1}{3} e^{-\frac{2 \gamma _a+3 \lambda _{aa}}{4 T}}
   \left(1+e^{\frac{\gamma _a}{T}}\right) \left(-1+e^{\frac{3 \lambda
   _{aa}}{2 T}}\right).
\end{eqnarray}
$\rho_{12}$ in equation (\ref{19}) is a special case of so called $X$-state \cite{xstate}. The concurrence $C(\rho)$
of such a density matrix has the following form \cite{form}:
\begin{equation}
C(\rho)=\frac{2}{Z}max(|y|-\sqrt{u v}, 0). \label{21}
\end{equation}
Finally, we consider transcendental equations (\ref{8a}) and
(\ref{8b}) by taking into account the values of variational
parameters: $\lambda_{aa}=J_{aa}$, $\gamma_a=2J_{ab}m_b+H$,
$\gamma_b=4J_{ab}m_a+H$, and, therefore, one can use these
parameters to calculate $C(\rho)$. First, we
study the behavior of $C(\rho)$
at $H=0$.
The
temperature dependence of $C(\rho)$ is shown in figure \ref{C_T_0}.

\begin{figurehere}
\begin{center}
\includegraphics[width=5.8cm]{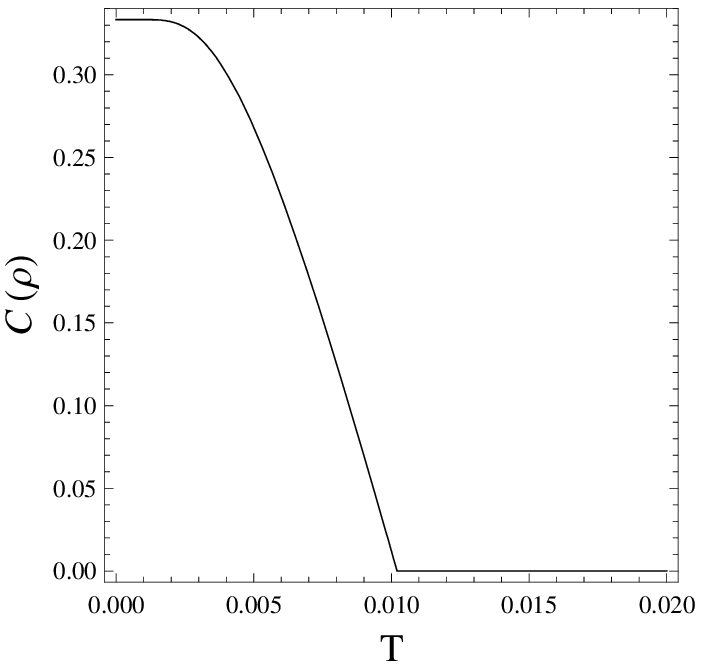}
\caption {\small{Concurrence $C(\rho)$ versus temperature field for
$J_{aa}=1$, $\alpha=0.025$ and $H=0$ \label{C_T_0}}}
\end{center}
\end{figurehere}

Notice, the "triangle-in-triangle" system can display
(bipartite)
entanglement
described by concurrence
even in the absence of external magnetic field. It is important to
mention that this result does not contradict to the well-known
fact that there is no
(bipartite)
entanglement,
measured by concurrence
in
isotropic
three-qubit Heisenberg $XXX$ model in zero magnetic field ($H=0$) \cite{wang}.
Indeed this effect is
due to the existence of Ising-type interaction replaced by
effective field $\gamma_a=2J_{ab}m_b+H$ acting upon $a$-spins. The
latter, in addition to $H$ contains
another quantity having
meaning of magnetic
($2J_{ab}m_b$), which is non-zero at $H=0$.

Another important observation is that threshold temperature at
which entanglement $C(\rho)$ disappear is identical to the
critical temperature $T_c$ of second order phase transition
between ordered and disordered phases described earlier in section
\ref{heat}. This implies that the concurrence vanishes precisely
at $T_c$, the same temperature of specific heat discontinuity.
This is the consequence of the fact that at $T_c$ the system
undergoes order-disorder phase transition and the second term in
$\gamma_a$ vanishes, too ($m_b=0$, when $H=0$ and $T\geq T_c$).
This factor implies the strong
relationship between magnetic and entanglement properties of the
system.
In figure \ref{C_T_3} we present the three dimensional plot of the
concurrence as a function of the temperature and external magnetic
field. We will discuss some of these features in behavior of
concurrence $C(\rho)$
for studying magnetic and entanglement thermal properties in section
\ref{identity}.

\begin{figurehere}
\begin{center}
\includegraphics[width=8cm]{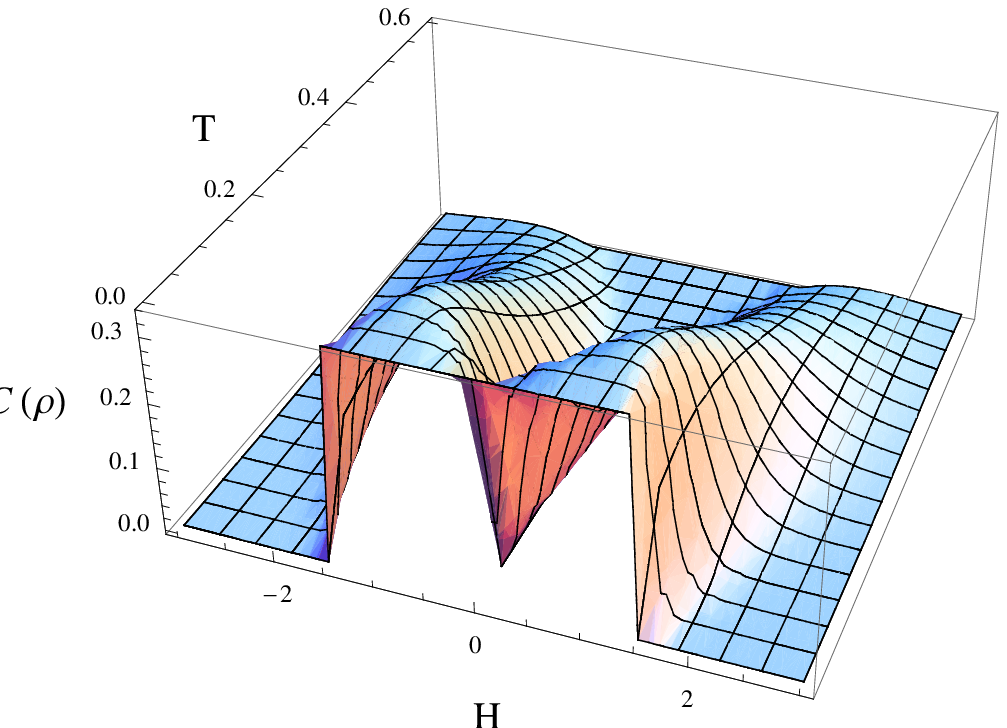}
\caption { \small {Concurrence $C(\rho)$ versus temperature $T$ and
external magnetic field $H$ for $J_{aa}=1$, $\alpha=0.025$.}}
\label{C_T_3}
\end{center}
\end{figurehere}

\section{Common features of magnetic properties and
entanglement }\label{identity}
\subsection{Finite temperatures \label{nonzero}}

To the best of our knowledge, the common thermal features in
entanglement and magnetic properties in frustrated systems have
not yet been reported, neither theoretically, nor experimentally.
In this section
%
%
we discuss some similarities of magnetic statistical properties
and quantum entanglement.

First, we consider
for general $J_{aa}$ and $H$ the
susceptibility
(\ref{13}) as a statistical characteristic. Figure \ref{for_J} (a)
shows the density distribution of susceptibility reduced per one
$a$-site as a function of the coupling constant $J_{aa}$ and the
external field $H$, at a relatively high temperature $T=0.1$, which
is higher than $T_c$. The white stripes on the figure correspond to
peaks of the susceptibility. These stripes have a certain finite
width due to nonzero temperature. For consistency in figure
\ref{for_J} (b) a similar plot of concurrence density is shown for
the same values of $J_{aa}-H$ parameters.
The existence of entanglement in the infinite $XXX$ Heisenberg chains of spins-$1/2$  and spins-$1$ was pointed in \cite{witness}. Here weak probe fields have been aligned along three orthogonal directions ($x$, $y$ and $z$), supposing that magnetic susceptibility is equal in all these directions ($\chi_x=\chi_y=\chi_z$) and using the fact that $\chi_x+\chi_y+\chi_z$ is an entanglement witness.
In our case there is a magnetic field aligned in the z-direction only, therefore $\chi_z\equiv\chi_a$. And we show that the behavior of susceptibility $\chi_z$ is similar to that of
bipartite entanglement.
Indeed, comparison of figures \ref{for_J}a and \ref{for_J}b
shows that the general behavior of the statistical and entanglement
properties, such as susceptibility ($\chi_a$) and concurrence
($C(\rho)$), coincide. Our calculations show that the values of
variables for the maximum (peak) in magnetic susceptibility
correspond to the critical values on the $J_{aa}-H$ diagram at which
the quantum coherence disappear and concurrence vanishes
($C(\rho)=0$).
However, this picture for the Ising-Heisenberg model
on the TKL lattice can applied only for antiferromagnetic coupling
$J_{aa}>0$, while for ferromagnetic coupling $J_{aa}<0$ the system
always remains disentangled.


\begin{center}
\begin{figurehere}
\begin{tabular}{ c c }
\small(a)  &  \small(b) \\
\begin{figurehere}
\includegraphics[width=7cm]{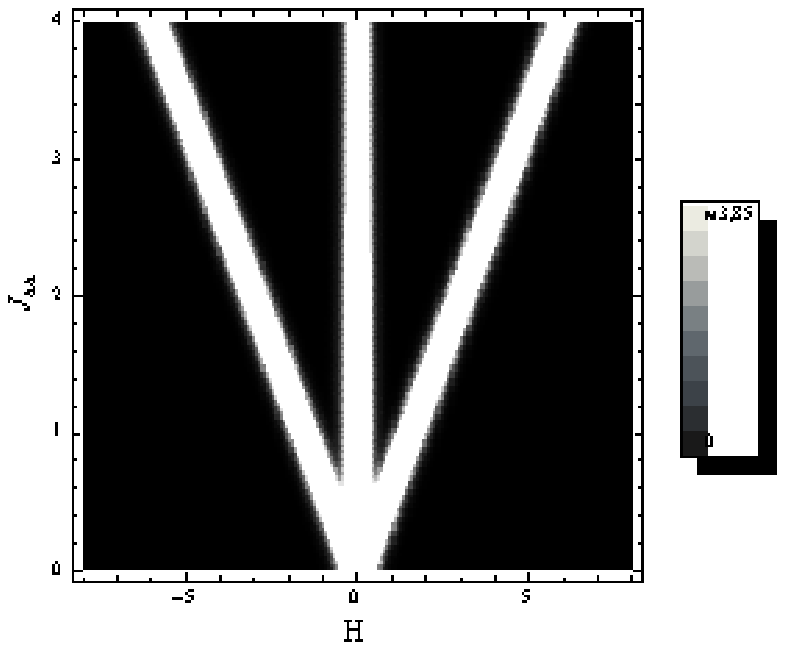}
\end{figurehere} &
\begin{figurehere}
\includegraphics[width=7cm]{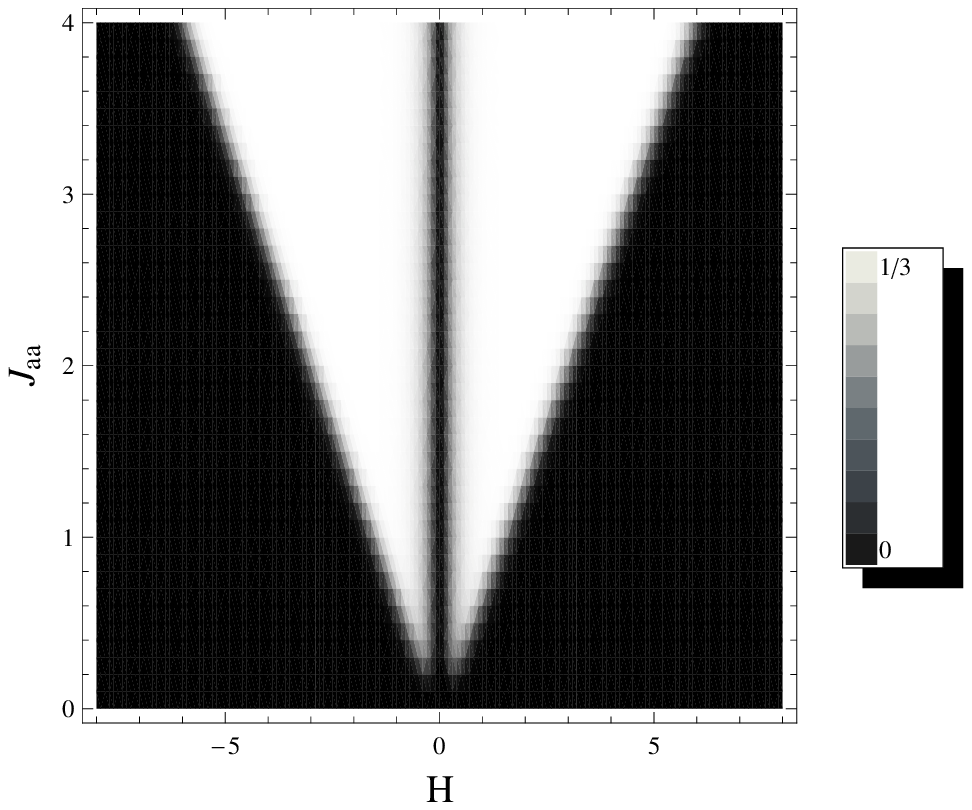}
\end{figurehere}  \\
\end{tabular}
\caption {\small{Density plot for (a) susceptibility $\chi_a$ and
(b) concurrence $C(\rho)$ versus magnetic field $H$ and coupling
constant $J_{aa}$ at $\alpha=0.025$ and $T=0.1$. \label{for_J}}}
\end{figurehere}
\end{center}

For further comparison we show for various temperatures in figure
\ref{spec_C} (a) and (b) the corresponding dependencies of the
concurrence
and heat capacity on magnetic field $H$.

\begin{center}
\begin{figurehere}
\begin{tabular}{ c c }
\small(a)  &  \small(b) \\
\begin{figurehere}
\includegraphics[width=7cm]{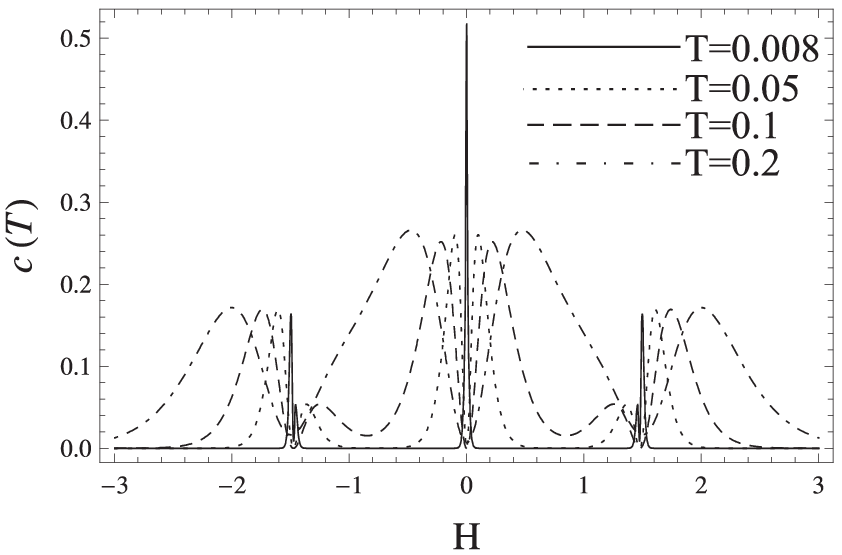}
\end{figurehere} &
\begin{figurehere}
\includegraphics[width=7cm]{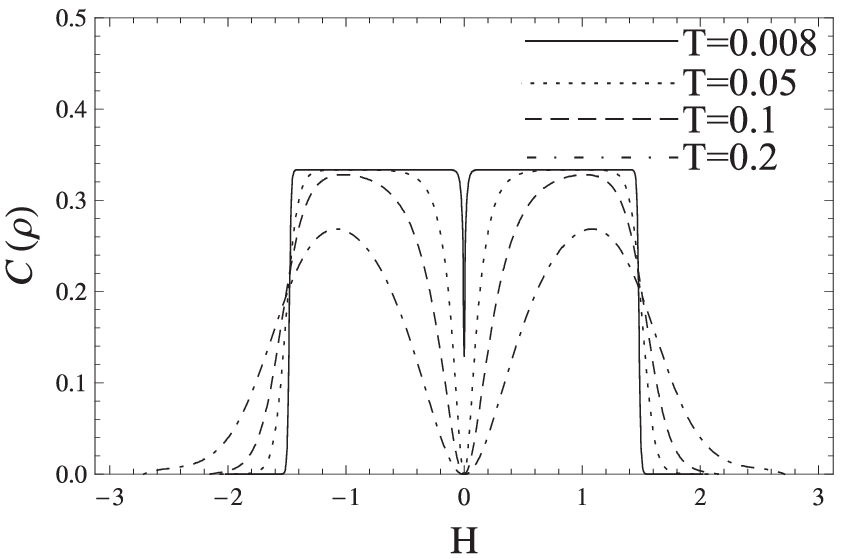}
\end{figurehere}  \\
\end{tabular}
\caption {\small{(a) Specific heat $c(T)$ and (b) concurrence
$C(\rho)$ versus external magnetic field $H$ for $J_{aa}=1$,
$\alpha=0.025$. \label{spec_C}}}
\end{figurehere}
\end{center}

In figure \ref{spec_C} (a) at relatively low temperatures the
specific heat exhibits six
peak structure located symmetrically with respect to the magnetic field ($H=0$).
As temperature increases, the middle peaks
(on both sides of the $H =0$)
splits and merge with the left
and right peaks in the neighbor areas, near $H=0$.
At higher temperatures, the two other
peaks on both sides of $H=0$ also merge in one.
As temperature increases there remain only two peaks,
i.e., the sharp peak structure gradually disappear.

At low temperatures close to ($T>T_c$), the two most distant peaks
from the $H=0$ (on each side of $H$) in figure \ref{spec_C} (a) are
approaching closer to each other, but do not merge in one.
Meanwhile, the closest ones to the $H=0$ peak (on either side of
$H$) becomes narrower and approach closer to the origin, $H=0$. For
$T\lesssim T_c$, some features are resulting from the effective
Ising field: the local minimum of the curve $c(T)$ at $H=0$ becomes
non zero, $c(T)|_{H=0}\neq 0$. With further decreasing temperature,
the heat capacity in the vicinity of $H=0$ displays one narrow peak
(whereas in a normal three-qubit Heisenberg model there are two
symmetrical peaks and $c(T)|_{H=0}= 0$). Nevertheless, the described
extreme effects do not affect the behavior of the entanglement
(concurrence), which is a purely quantum feature: the curve is
smooth enough for all $T>T_c$. However, $C(\rho)$ behaves as a
step-like function by approaching closer to critical temperature
$T_c$. At temperatures below the critical value, a local dip minimum
is non zero, $C(T)|_{H=0}\neq 0$ (see also figure \ref{C_T_0}). This
dip disappears at $T\approx 0.002$ by forming a single flat plateau
at $T=0$.

Our effective field results for coupled spins controlled by external
magnetic field provide understanding of the magnetic ground state
properties in "triangles-in-triangles" Cu$_9$X$_2$(cpa)$_6$
Kagom\'{e} series in \cite{Maruti}, which are similar to the one
dimensional systems. This simple approach can also explain several
interesting properties, such as
double peaks of the specific heat,
different competing (magnetic) orders,
a 1/3 magnetization plateau and susceptibility peaks
for the pulse field reported for classical and quantum Kagom\'{e} lattice
magnets in \cite{Maegawa}.
In the end of the section, we emphasize that
although introduced (effective) self-consistent $\gamma_a$ and $\gamma_b$
fields break the symmetry against $H=0$, this does not act upon
concurrence and specific heat.

\subsection{Zero temperature entanglement and modulated phases \label{ground}}
In this section the magnetization and entanglement properties of
$a$-sublattice are considered at zero temperature using
variational mean-field approximation. In figure~\ref{zero}(a) a
phase diagram of constant magnetization is shown for
$a$-sublattice. This diagram differentiates the following phases:
Phase $\mathbf{I}$ corresponds to the single $a$-site
magnetization $m_a=1/6$, when spins in $a$-sublattice are in
$\uparrow\uparrow\downarrow$ configuration; Phase $\mathbf{II}$
corresponds to $\downarrow\downarrow\uparrow$ configuration with
the single $a$-site magnetization $m_a=-1/6$. These phases exist
only for the antiferromagnetic coupling, $J_{aa}>0$. For the
ferromagnetic case ($J_{aa}<0$) in $\mathbf{III}$ and
$\mathbf{IV}$ regions we get spin saturation, with maximum
$m_a=1/2$ ($\downarrow\downarrow\downarrow$) and minimum
$m_a=-1/2$ ($\uparrow\uparrow\uparrow$) magnetization per atom
respectively.

Phase $\mathbf{I}$ contains the two-fold degenerate states
$|\psi_5\rangle$ and $|\psi_6\rangle$, while Phase
$\mathbf{II}$-the two-fold degenerate states $|\psi_2\rangle$ and
$|\psi_3\rangle$. By constructing the reduced density matrix in
Phases $\mathbf{I}$ and $\mathbf{II}$, one can find these phases
in maximum entangled state, $C(\rho)=1/3$. Phases $\mathbf{III}$
and $\mathbf{IV}$ correspond to $|\psi_1\rangle$ and
$|\psi_8\rangle$ states respectively. These phases are
disentangled, $C(\rho)=0$. In figure \ref{zero}(b) the concurrence
density distribution is shown versus coupling constant ($J_{aa}$)
and magnetic field ($H$) at zero temperature. The area of non-zero
entanglement coincides with the phase $\mathbf{I}+\mathbf{II}$,
where $|m_a|=1/6$, while the one with zero entanglement
($C(\rho)=0$) corresponds to the phase $\mathbf{III}+\mathbf{IV}$
with $|m_a|=1/2$.

Notice, the plateau
behavior in magnetization corresponds to constant entanglement
values in corresponding density plots. The plateau
in magnetization at $|m_a|=1/6$ corresponds to maximum
entanglement value, $C(\rho)=1/3$, where the saturated phase at
$|m_a|=1/2$ is disentangled, $C(\rho)=0$. This descriptive picture
is also available at the non-zero temperature. At relatively low
temperatures the plateau  of magnetization at $|m_a|=1/6$ and entanglement coincide, except
the narrow region in the vicinity of $H=0$ border. By decreasing
the temperature the middle stripe in figure \ref{for_J}(b) narrows
and gradually disappears (see figure \ref{zero}(b) and also
section~\ref{nonzero}). This trend becomes apparent by comparison of
figures \ref{for_J}(b) and \ref{plateau}.
The latter represents the density distribution of $m_a$
magnetization at considerably low temperature $T=0.1$.
In figure \ref{plateau} the grey areas describe the plateau at
$|m_a|=1/6$, while black and white regions correspond to saturated
states, $|m_a|=1/2$. White regions in figure \ref{plateau}
correspond to the plateau behavior in the concurrence. As the
temperature increases the borders between distinct (different)
phases are gradually smeared out.
Summarizing, the structure of each of the Heisenberg trimers has the crucial impact on the phase diagram in figure~12(a): the geometrical structure of a-sublattice is responsible for the frustration effects arising in the antiferromagnetic Heisenberg model (geometrical frustraion). This leads to above mentioned ground states with definite values of concurrence in figure 12(b). The ground state concurrence arises on magnetization plateaus at $|m_a|=1/6$ (see figure 2(b) for non-zero temperatures), which is a consequence of strong geometrical frustration of a-sublattice. While the octahedron environment (b-sites) are responsible for the effective field only.
\begin{center}
\begin{figurehere}
\begin{tabular}{ c c }
\small(a)  &  \small(b) \\
\begin{figurehere}
\includegraphics[width=5.7cm]{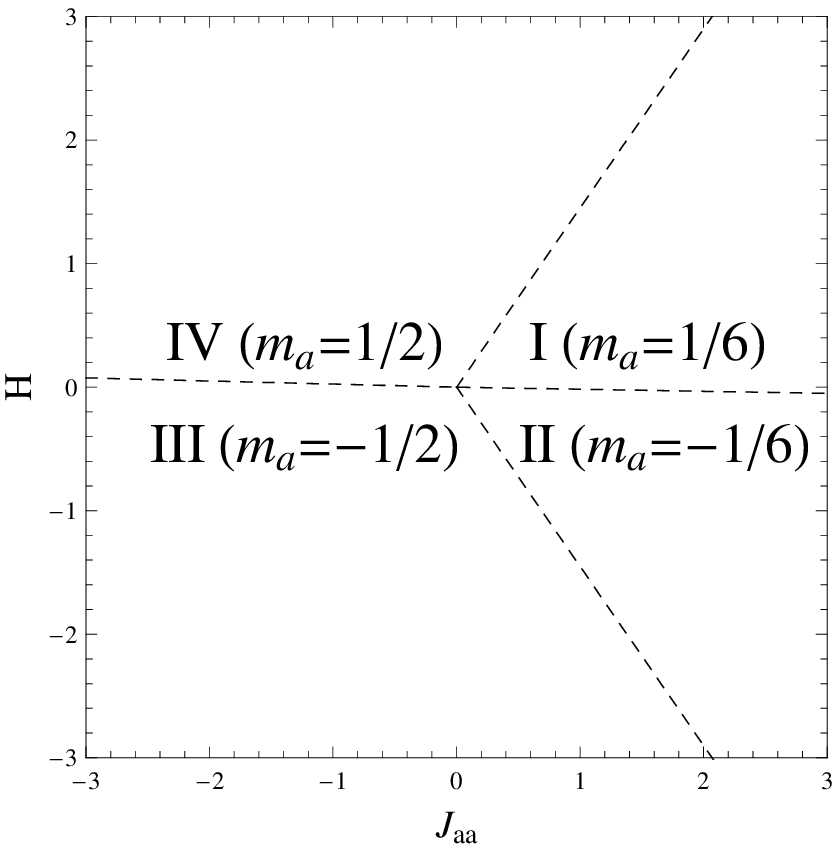}
\end{figurehere} &
\begin{figurehere}
\includegraphics[width=7cm]{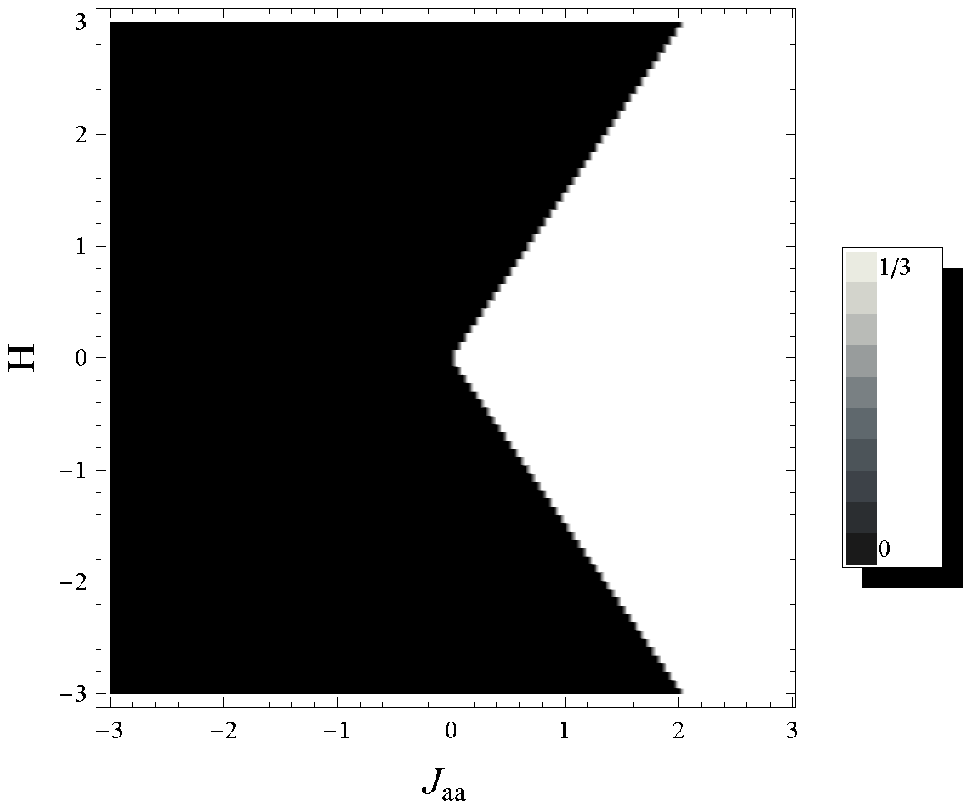}
\end{figurehere}  \\
\end{tabular}
\caption {\small{(a) Phase diagram of $a$-sublattice for
$|\alpha|=0.025$ and (b) density plot of concurrence $C(\rho)$
versus magnetic field $H$ and coupling constant $J_{aa}$ for
$|\alpha|=0.025$ at zero temperature. \label{zero}}}
\end{figurehere}
\end{center}

\begin{figurehere}
\begin{center}
\includegraphics[width=7cm]{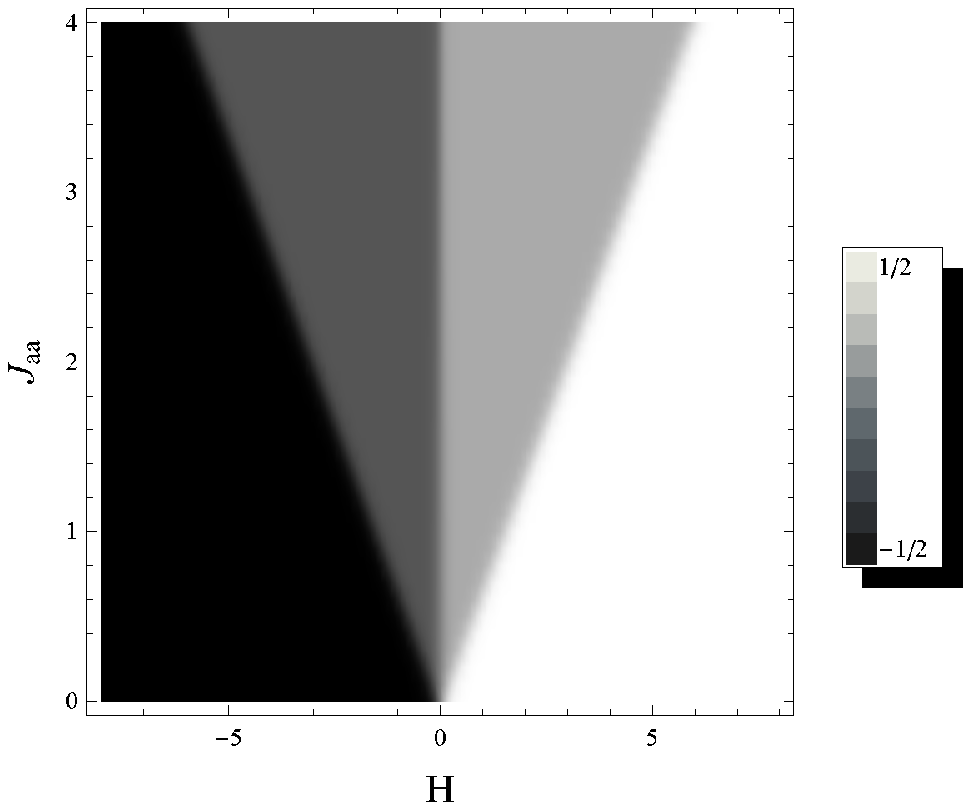}
\caption {\small{Density plot of magnetization $m_a$ versus magnetic
field $H$ and coupling constant $J_{aa}$ for $\alpha=0.025$ and
$T=0.1$. \label{plateau}}}
\end{center}
\end{figurehere}

\section{Conclusion \label{concl}}

In this paper we find strong
correlations between magnetic properties and quantum entanglement in
spin-$\frac{1}{2}$ Ising-Heisenberg model on triangulated Kagom\'{e}
lattice, which has been proposed to understand a frustrated
magnetism of the series of $\mathrm{Cu_9X_2(cpa)_6\cdot n H_2O}$
polymeric coordination compounds. The ratio
$\alpha=J_{ab}/J_{aa}=0.025$ ($J_{aa}$ labels intra-trimer
Heisenberg, while $J_{ab}$ monomer-trimer Ising interactions) is
considered, which guaranties experimental realization for suitable
theoretical treatment. We adopted variational mean-field like
treatment (based on Gibbs-Bogoliubov inequality) of separate
clusters in effective interconnected fields of two types (consisting
of Heisenberg $a$ trimers and Ising-type $b$ monomers).
Each of these fields taken separately describes not only
corresponding (a- or b- type) spins, but the whole system.

The calculated magnetic and thermodynamic properties of the model
display the (smooth) second order phase
transition from ordered into disordered phase driven by
temperature and magnetic field. The thermal entanglement
properties of $a$-trimers are
the central in the aforementioned approximation. Since there are
some open questions in the definition of multipartite entanglement
(for example, see \cite{multi, multi1}) we used concurrence as a
computable measure of bipartite entanglement for the trimeric
units in terms of $XXX$ Heisenberg model in self-consistent
magnetic field applied to $a$ subsystem.
Due to the classical character of Ising-type interactions
the states of two neighboring Heisenberg trimers are separable. Using
this fact we studied the entanglement of the each "$a$
subdivisions" of the cluster individually in effective Ising-type
field.Our results show that entanglement does not vanish in zero
external field as it happens for the mean field treatment of the
isotropic $XXX$ Heisenberg model on triangulated lattice.
The geometrical structure of the lattice is responsible for the frustration effects arising in the antiferromagnetic Heisenberg model (geometrical frustration). It leads to existence of magnetization plateaus at $|m_a|=1/6$ and concurrence at $C(\rho)=1/3$. Besides the effective field in mean-field solution also depends on the lattice structure (via the number of $b$-sites in the cluster).

In addition, the entangled-disentangled phases in concurrence and
order-disordered phases in quantum phase transitions share many
common features. The threshold temperature for concurrence is
identical to critical temperature of second order phase
transition. Besides the entanglement and thermodynamic properties
exhibit also common (plateau and peak) behavior in magnetization,
susceptibility and concurrence. Moreover, we
found that for antiferromagnetic interaction the magnetic
susceptibility peaks coincide with the corresponding phase
boundaries at which the entanglement vanishes.
However, this does not take place for the classical ferromagnetic
interaction. This fact allows one to notice a quite visible correlation for the boundaries between various
phases for entanglement, susceptibility
and magnetization densities as a fingerprints of corresponding
quantum phase transitions. Thus, the density diagrams in the
presence of magnetic field can be considered as a useful tools to
detect important relationships between
entanglement-disentanglement transitions in concurrence and
corresponding order-disorder quantum phase transitions in quantum
magnetism.

\section{Acknowledgements}

This work was supported by the PS-1981 ANSEF and ECSP-09-08-sasp
NFSAT research grants.

\end{document}